\documentstyle[12pt,epsfig]{article}
\date{}
\begin{document}

{\large \sf
\title{
{\normalsize
\begin{flushright}
\end{flushright}}
{\Huge \sf New Insights to Old Problems}\thanks{{\normalsize \sf
Based on an invited talk given at the 2006 APS Meeting as the
first presentation in the Session on 50 Years Since the Discovery
of Parity Nonconservation in the Weak Interaction {\rm I}, April
22, 2006 (Appendix A)}}}

{\large \sf
\author{
{\large \sf T. D. Lee}\\
{\normalsize \it Pupin Physics Laboratory, Columbia University, New York, NY, USA}\\
{\normalsize \it and}\\
{\normalsize \it China Center of Advanced Science and Technology
(CCAST), Beijing, China}}
\maketitle

\begin{abstract}

{\normalsize \sf From the history of the $\theta$-$\tau$ puzzle
and the discovery of parity non-conservation in 1956, we review
the current status of discrete symmetry violations in the weak
interaction. Possible origin of these symmetry violations are
discussed.}

\end{abstract}

\vspace{1cm}

{\normalsize \sf PACS{:~~11.30.Er,~~12.15.Ff,~~14.60.Pq}}

\vspace{1cm}

{\normalsize \sf Key words: history of discovery of parity
violation, mixing angles, origin of symmetry

~~~~~~~~~~~~~~~violation, scalar field and properties of vacuum}

\newpage

\section*{\Large \sf  1. Symmetry Violations: The Discovery}
\setcounter{section}{1} \setcounter{equation}{0}

Almost exactly 50 years ago, I had an important conversation with
my dear friend and colleague C. S. Wu. This conversation was
critical for setting in motion the events that led to the
experimental discovery of parity nonconservation in $\beta$-decay
by Wu, et al.[1]. In the words of C. S. Wu[2]:

"$\cdots$ One day in the early Spring of 1956, Professor T. D. Lee
came up to my little office on the thirteenth floor of Pupin
Physical Laboratories. He explained to me, first, the
$\tau$-$\theta$ puzzle. If the answer to the $\tau$-$\theta$
puzzle is violation of parity--he went on--then the violation
should also be observed in the space distribution of the
beta-decay of polarized nuclei: one must measure the pseudo-scalar
quantity $<\sigma\cdot {\bf p}>$ where ${\bf p}$ is the electron
momentum and $\sigma$ the spin of the nucleus.

$\cdots$ Following Professor Lee's visit, I began to think things
through. This was a golden opportunity for a beta-decay physicist
to perform a crucial test, and how could I let it pass? $\cdots$
That Spring, my husband, Chia-Liu Yuan, and I had planned to
attend a conference in Geneva and then proceed to the Far East.
Both of us had left China in 1936, exactly twenty years earlier.
Our passages were booked on the Queen Elizabeth before I suddenly
realized that I had to do the experiment immediately, before the
rest of the Physics Community recognized the importance of this
experiment and did it first. So I asked Chia-Liu to let me stay
and go without me.

$\cdots$ As soon as the Spring semester ended in the last part of
May, I started work in earnest in preparing for the experiment.
$\cdots$ In the middle of September, I finally went to Washington,
D. C. for my first meeting with Dr. Ambler. $\cdots$ Between
experimental runs in Washington, I had to dash back to Columbia
for teaching and other research activities. On Christmas eve, I
returned to New York on the last train; the airport was closed
because of heavy snow. There I told Professor Lee that the
observed asymmetry was reproducible and huge. The asymmetry
parameter was nearly -1. Professor Lee said that this was very
good. This result is just what one should expect for a
two-component theory of the neutrino."

A few days later in January 1957, my other Columbia University
colleagues Leon Lederman and Richard Garwin followed with their
experiment on parity nonconservation in $\pi-\mu-e$ decay[3] (as
will be discussed in the second presentation of today's session by
Leon Lederman himself). An explosion of hundreds of other
experiments on parity (P) and charge conjugation (C) followed
quickly in many physics laboratories all over the world. Their
results answered the question that I raised with C. N. Yang in our
1956 paper[4] on "Question of Parity Conservation in Weak
Interactions".

Columbia University was established in 1754 when America was still
a British colony. In 1954 on the occasion of the 200th anniversary
of Columbia University, the great theoretical physicist W. Pauli
was invited to give a special Bicentennial Lecture at Columbia.
His topic was CPT Theorem[5]. As a member of the Columbia physics
department, I was in the audience. After his speech, Pauli, I. I.
Rabi, C. S. Wu and I had dinner together. Both Rabi and Wu were
Pauli's good friends. Pauli's lecture impressed me. By pure
deduction, Pauli proved that for a Lorenz invariant local field
theory, the combined invariance of CPT naturally follows. Of
course, at that time everyone assumes that C, P and T are
separately conserved. That one can derive CPT invariance based on
locality and Lorenz invariance is interesting, but more as an
intellectual curiosity. Two years later in 1956, when
$\theta-\tau$ puzzle led to the question of parity violation,
Pauli's CPT Theorem took on a new significance. If P were
violated, at least one other discrete symmetry must also be
violated: C or T, or both?

In general, it is difficult to observe T violation (because of CPT
Theorem, the same difficulty applies also to CP violation). In the
same year, 1956, Leon Lederman and his group discovered[6] the
long lived neutral kaon $K^0_L$ (called $\theta^0_2$ then). This
discovery was heralded at that time as the proof of C invariance,
based on the theoretical analysis of Gell-Mann and Pais[7]. If
that were the case, with C conservation and only P and T
violations, neither $\pi-\mu-e$ decay nor $\beta$-decay could show
any sizable observable parity violation effects.

Thus, it became necessary for me to examine the theoretical
analysis of Gell-Mann and Pais. With only CPT invariance, the
paper of Lee, Oehme and Yang[8] proved that the existence of a
long-lived neutral kaon is the consequence of only CPT invariance.
It does not imply C invariance. Assuming CPT invariance the same
Lee-Oehme-Yang paper questioned the validity of T and, therefore,
also CP invariance. The analysis of interplay between $K^0_S$,
$K^0_L$ and the strangeness eigenstates (called $K_1$ and $K_2$
then) was given in the same paper. This provided the theoretical
basis for the 1964 experimental discovery of CP and T violations
by Cronin, Fitch and their co-workers[9].

\section*{\Large \sf  2. Present Status: The Complexity}
\setcounter{section}{2} \setcounter{equation}{0}

The hadronic weak current converts the three charge $2e/3$ quark
mass eigenstates $u,~c,~t$ into three charge $-e/3$ states
$d',~s'~,b'$, which are in turn linear superpositions of the mass
eigenstates $d,~s,~b$ through the $SU(3)$ CKM matrix $U_h$[10]:
\begin{eqnarray*}\label{e1}
~~~~~~~~~~~~~~~~\left\{
\begin{array}{c}
d'\\
s'\\
b'
\end{array}
\right\} =U_h
\left\{
\begin{array}{c}
d\\
s\\
b
\end{array}
\right\}.~~~~~~~~~~~~~~~~~~~~~~~~~~~~~~~~~~~~~~~~~~~~~~~~~~~(1)
\end{eqnarray*}
Any $SU(3)$ matrix depends on 8 real parameters. Between the
initial $(u,~c,~t)$ states there are 2 arbitrary relative phases;
likewise, there are 2 additional arbitrary relative phases between
the three final states $(d,~s,~b)$. It is customary[11] to choose
the remaining $8-2^2=4$ parameters by four angles,
$\theta_{12},~\theta_{23},~\theta_{13}$ and a phase angle
$\delta$:
\begin{eqnarray*}\label{e2}
~~~\left\{
\begin{array}{ccc}
c_{12}c_{13}&s_{12}c_{13}&s_{13}e^{-i\delta}\\
-s_{12}c_{23}-c_{12}s_{23}s_{13}e^{i\delta}&c_{12}c_{23}-s_{12}s_{23}s_{13}e^{i\delta}&s_{23}c_{13}\\
s_{12}s_{23}-c_{12}c_{23}s_{13}e^{i\delta}&-c_{12}s_{23}-s_{12}c_{23}s_{13}e^{i\delta}&c_{23}c_{13}
\end{array}
\right\}.~~~~~~~~~~~~~~~~~(2)
\end{eqnarray*}
with $c_{ij}=\cos\theta_{ij}$ and $s_{ij}=\sin\theta_{ij}$ for the
generation labels $i,j=1,~2,~3$. The approximate experimental
determination of the CKM matrix $U_h$ can then be written as
\begin{eqnarray*}\label{e3}
~~~~~~~~U_h\simeq \left\{
\begin{array}{ccc}
0.974&0.227&(2-3i)10^{-3}\\
-0.227&0.973&0.04\\
(7-3i)10^{-3}&-0.04&0.999
\end{array}
\right\}.~~~~~~~~~~~~~~~~~~~~~~~~~~(3)
\end{eqnarray*}

Likewise, the leptonic weak current connects the three charge
leptons $e^-,~\mu^-$ and $\tau^-$ into three neutral states
$\nu_e,~\nu_\mu$ and $\nu_\tau$; which are in turn coherent
superpositions of the three mass eigenstates $\nu_1,~\nu_2$ and
$\nu_3$:
\begin{eqnarray*}\label{e4}
~~~~~~~~~~~~~~~~\left\{
\begin{array}{c}
\nu_e\\
\nu_\mu\\
\nu_\tau
\end{array}
\right\} =U_l \left\{
\begin{array}{c}
\nu_1\\
\nu_2\\
\nu_3
\end{array}
\right\}~~~~~~~~~~~~~~~~~~~~~~~~~~~~~~~~~~~~~~~~~~~~~~~~~~~(4)
\end{eqnarray*}
where $U_l$ can be cast into the same form (2) and given
approximately by[11]
\begin{eqnarray*}\label{e5}
~~~~~~~~U_l\simeq \left\{
\begin{array}{ccc}
0.84&0.56&s_{13}e^{-i\delta}\\
-0.4-0.6s_{13}e^{i\delta}&0.6-0.4s_{13}e^{i\delta}&0.7\\
0.4-0.6s_{13}e^{i\delta}&-0.6-0.4s_{13}e^{i\delta}&0.7
\end{array}
\right\}~~~~~~~~~~~~~~~~~(5)
\end{eqnarray*}
with
\begin{eqnarray*}\label{e6}
~~~~~~~~~~~~~~~~~~~s_{13}^2=\sin^2\theta_{13}=0.9 \left.
\begin{array}{cc}
+2.3\\
-0.9
\end{array}
\right. \times 10^{-2}~~~~~~~~~~~~~~~~~~~~~~~~~~~~~~~(6)
\end{eqnarray*}
and $\delta$ unknown. The three neutrino mass states form a
doublet $\nu_1,~\nu_2$, with
$$
|m^2_2-m^2_1|=7.92(1\pm 0.09)\cdot 10^{-5}{\sf ev}^2\eqno(7)
$$
and a third singlet $\nu_3$ with mass $m_3$ given by
\begin{eqnarray*}\label{e8}
~~~~~~~~~~~\bigg |~m^2_3-\frac{m^2_1+m^2_2}{2}~\bigg |=2.4\times
(1 \left.
\begin{array}{cc}
+0.21\\
-0.26
\end{array}
\right.
 )\times 10^{-3}{\sf ev}^2.~~~~~~~~~~~~~~~~~~~(8)
\end{eqnarray*}
The signs of these mass differences are not known. As a
convention, $\nu_1$ refers to the lighter member of the doublet.

Yes, parity is violated. CP is violated and so are C and T. The
questions raised fifty years ago are all answered. Neutrinos are
not two-component and leptons have three generations, like the
quarks. These two matrices CKM and neutrino mapping represent the
fruition of the tremendous 50 year-effort of nuclear and particle
physics. The question is: can we read this Rosetta
stone\footnote{{\normalsize \sf Rosetta stone is an ancient stone
found at a small village named Rosetta in Egypt in 1799. The text
carved on the stone in 196 B.C. is in Egyptian and Greek, using
three different scripts. Many people worked for over several
hundred years to decipher the text written in hieroglyph.}}?

\newpage

\section*{\Large \sf  3. Through the Looking Glass[12]}
\setcounter{section}{3} \setcounter{equation}{0}

\noindent (i) Can $U_h\rightarrow U_l$ at high energy?

Assume that the only differences between quarks and leptons are
QCD and their masses. Since QCD is an asymptotically free theory,
at high energy, QCD forces should be asymptotically free. Thus,
except for the quark-lepton mass differences, the CKM matrix $U_h$
would become the same as the neutrino mapping matrix $U_l$ at high
energy. This hypothesis can be tested either by high energy
experiments, or by doing a lattice QCD calculation and examining
the change of CKM matrix elements from their low energy values
given by (3) to their high energy values given by (5). In the
following, we shall assume that this is indeed the case: the CKM
matrix $U_h$ does approach the neutrino mapping matrix $U_l$ at
very high energy. Thus, the task of understanding the mystery of
two matrices CKM and $\nu$-mapping is reduced to that of one
matrix, the $\nu$-mapping matrix $U_l$.\\

\noindent (ii) An approximate geometric representation of $U_l$

\begin{figure}[h]
 \centerline{
\epsfig{file=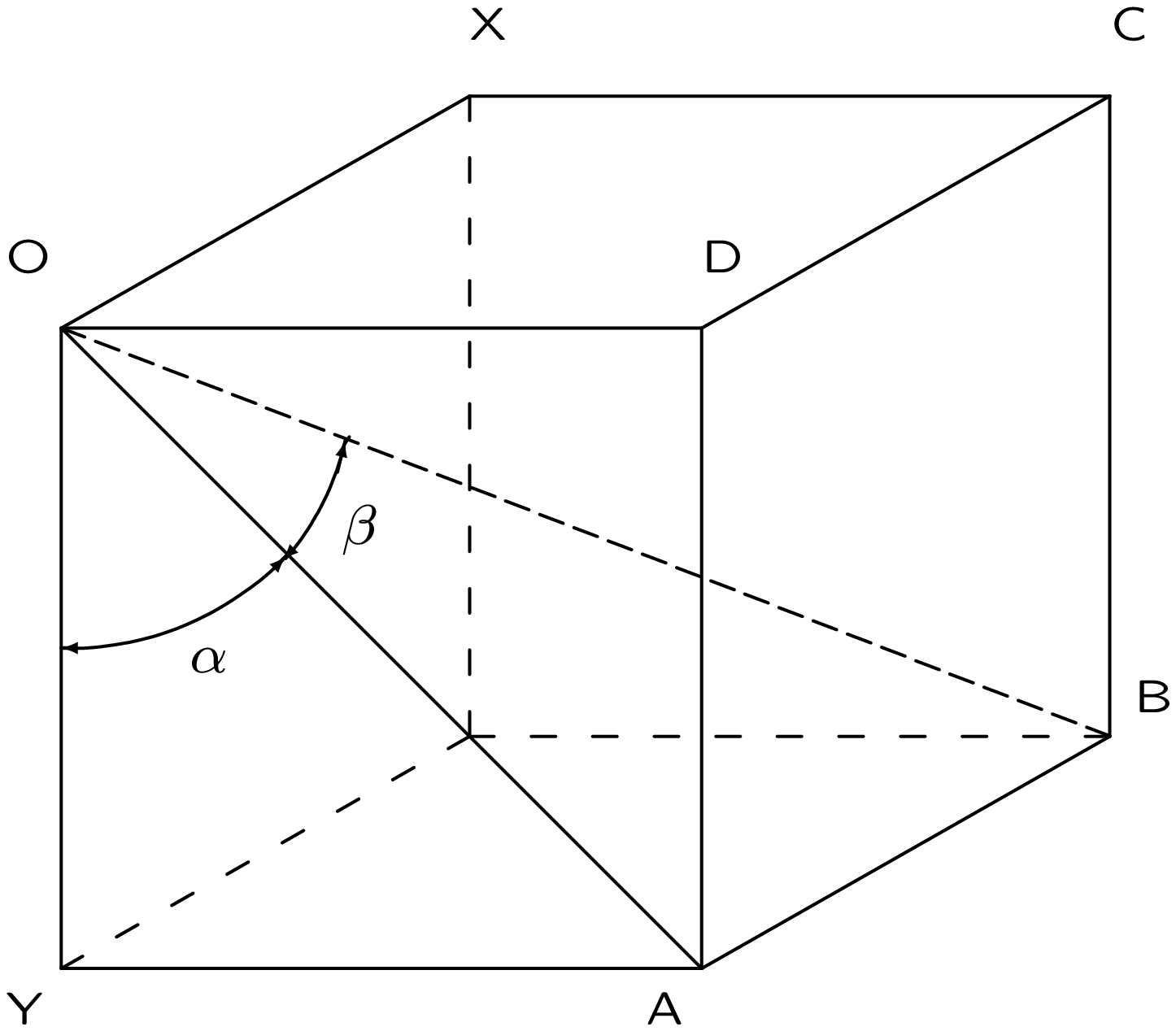, width=8cm, height=6.8cm}}
\vspace{.2cm}
 \centerline{{\normalsize \sf Fig. 1~  A Geometric representation of the
 neutrino-mapping matrix
 (see Eq. (12)).}}
\end{figure}

We assume that the $\nu$-mapping matrix elements are all
determined by vacuum expectation values of a set of spin $0$
fields. In that case, these vacuum expectation values can acquire
a geometric shape[13]. As an approximation, we may set $s_{13}
e^{i\delta}\simeq 0$, in accordance with (6); this reduces $U_l$
to a standard rigid body rotation matrix in three dimensions.

Consider the cube in Figure 1, in which we designate the seven
visible corners in the figure as O, X, Y and A, B, C, D. Regard
$\overline{OX},~ \overline{OY}$ and $\overline{OZ}=-\overline{OD}$
as the Cartesian coordinates along the state vectors representing
$\nu_e,~\nu_\mu$ and $\nu_\tau$. Let the angles $\alpha$ and
$\beta$ be
$$
\alpha=\angle YOA = 45^0 ~~~{\sf and}~~~\beta=\angle AOB =
\sin^{-1}\sqrt{\frac{1}{3}}.\eqno(9)
$$
A left-hand rotation of $\alpha=45^0$ along $\overline{OX}$ is
represented by the matrix
\begin{eqnarray*}\label{e10}
~~~~~~~~~~~~~~~~~~R_1= \left\{
\begin{array}{ccc}
1&0&0\\
0&\sqrt{\frac{1}{2}}&-\sqrt{\frac{1}{2}}\\
0&\sqrt{\frac{1}{2}}&\sqrt{\frac{1}{2}}
\end{array}
\right\},~~~~~~~~~~~~~~~~~~~~~~~~~~~~~~~~~~~~~~~~~~~~(10)
\end{eqnarray*}
which rotates $\overline{OY}$ and $\overline{OZ}$ to
$\overline{OY}'$ and $\overline{OZ}'$. A subsequent left-hand
rotation of $\beta=\sin^{-1}\sqrt{\frac{1}{3}}$ along
$\overline{OZ}'$ is represented by
\begin{eqnarray*}\label{e11}
~~~~~~~~~~~~~~~~~~~~R_2=\left\{
\begin{array}{ccc}
\sqrt{\frac{2}{3}}&-\sqrt{\frac{1}{3}}&0\\
\sqrt{\frac{1}{3}}&\sqrt{\frac{2}{3}}&0\\
0&0&1
\end{array}
\right\}.~~~~~~~~~~~~~~~~~~~~~~~~~~~~~~~~~~~~~~~~~~(11)
\end{eqnarray*}
Consider the transformation
\begin{eqnarray*}
\left\{
\begin{array}{c}
\nu_1\\
\nu_2\\
\nu_3
\end{array}
\right\} =R_2R_1 \left\{
\begin{array}{c}
\nu_e\\
\nu_\mu\\
\nu_\tau
\end{array}
\right\}
\end{eqnarray*}
i.e.,
\begin{eqnarray*}
\left\{
\begin{array}{c}
\nu_e\\
\nu_\mu\\
\nu_\tau
\end{array}
\right\} =\tilde{R}_1\tilde{R}_2\left\{
\begin{array}{c}
\nu_1\\
\nu_2\\
\nu_3
\end{array}
\right\}
\end{eqnarray*}
with $\sim$ denoting the transpose. From (10) and (11), we have
\begin{eqnarray*}\label{e12}
~~~~~~~~\tilde{R}_1\tilde{R}_2=\left\{
\begin{array}{ccc}
\sqrt{\frac{2}{3}}&\sqrt{\frac{1}{3}}&0\\
-\sqrt{\frac{1}{6}}&\sqrt{\frac{1}{3}}&\sqrt{\frac{1}{2}}\\
\sqrt{\frac{1}{6}}&-\sqrt{\frac{1}{3}}&\sqrt{\frac{1}{2}}
\end{array}
\right\} \cong \left\{
\begin{array}{ccc}
.8165&.5773&0\\
-.4082&.5773&.7071\\
.4082&-.5773&.7071
\end{array}
\right\}, ~~~~~~~(12)
\end{eqnarray*}
consistent with (5), our present knowledge of $U_l$ within the
experimental uncertainties of its matrix elements[11].\\

\noindent (iii) Origin of symmetry violation

We shall explore the attractive possibility that all symmetry
violations occur spontaneously; i.e., via the Higgs mechanism[14].
It follows then that the electro-weak gauge fields (i.e.,
$W^\pm_\mu, ~Z^0_\mu$ and $A^0_\mu$) are right-left symmetric, as
well as time reversal and CP symmetric. The simplest theory would
be the $SU(2)_L \times SU(2)_R \times U(1)$ model[15]; such a
model predicts the existence of additional heavier righthanded
$W^\pm_\mu(R)$ and $Z^0_\mu(R)$, which are as yet not
observed[11]. In this approach, all spin$\neq0$ fields are
symmetry conserving and, by themselves, are all of zero mass;
these include graviton in general relativity, $SU(3)$ color gauge
field in quantum chromodynamics, the electro-weak gauge field, as
well as the spin $\frac{1}{2}$ quark and lepton fields. Thus the
gauge fields are simple to comprehend, but the complexity of our
universe is rested in the spin $0$ fields.

There are good reasons to believe that the Higgs mass $m_H$ is not
too high, with $m_H<195 Gev$[11,16]. One reason for the
experimental difficulty of detecting Higgs could be that the
radial size $r_H$ of the Higgs is much larger than its Compton
wave length $m_H^{-1}$:
$$
r_H>m_H^{-1}.\eqno(13)
$$
Such an object would be hard to detect by searching for its
complex poles. A good example is the Cooper pair in
superconductivity[17]. Another example is the $\sigma$-model[18],
which is extremely valuable in describing the low energy
pion-nucleon physics; yet, the $\sigma$-pole itself is still not
discovered.

The search for Higgs is a deep and important subject; it has
relevance to several other major fields of physics, as we shall
discuss.

\newpage

\section*{\Large \sf  4. Importance of the Spin $0$ Field}
\setcounter{section}{4} \setcounter{equation}{0}

Recently there exists strong evidence[19] that the cosmological
constant $\Lambda$ is not only nonzero, but very large and most
likely not a  constant. Its energy density (also called dark
energy) is
$$
\rho_\Lambda \sim 3~-~7 \times 10^{-6} ~{\sf Gev/cm}^3\eqno(14)
$$
comparable to the critical energy density of our entire universe
$$
\rho_c \sim 1 \times 10^{-5}~ {\sf Gev/cm}^3.\eqno(15)
$$
Furthermore, $\Lambda$ corresponds to a negative pressure. What is
the origin of this negative pressure? Why should $\rho_\Lambda$
and $\rho_c$ be of the same order of magnitude?

At the end of last year, the American Institute of Physics
announced[20] that the top physics story of 2005 is the RHIC
discovery (at the Relativistic Heavy Ion Collider, Brookhaven
National Laboratory) of a strongly interacting quark-gluon plasma
(called sQGP), which behaves almost like a perfect fluid, with
very low viscosity. The search for such a new state of nuclear
matter has been a holy grail of high energy nuclear physicists
since early 1974[21], through intensive experimental efforts at
BEVALAC at Berkeley, AGS at Brookhaven and SPS at CERN. Now,
finally it is discovered at RHIC. However, the theoretical basis
and understanding of sQGP is still at the very beginning.
Undoubtedly, the start of LHC programs at CERN next year will be
critical to our future understanding. What is the nature of this
nearly perfect nuclear sQGP fluid? Does it depend only on quantum
chromodynamics? Does it also depend on other interactions?

In the following, I wish to present a unifying view: the negative
pressure of dark energy, the nearly perfect fluidity of sQGP and
the symmetry violations in the electroweak (as well as that in all
physics) are all due to a family of spin $0$ fields. The basic
mechanism is the same as that used in the MIT bag model[22] and
the Dirac model of muon[23].

Consider a scalar field $\phi$, which can be either a fundamental
field or a composite one, made of other fields, such as the
$\sigma$-model and Higgs field. Since by using the products of any
field, one can always construct a scalar  component, the physical
basis for the existence of such a scalar component $\phi$ is quite
general. The vacuum is a scalar, hence $\phi$ is coupled to the
vacuum and therefore, to the inertia of any physical particle. In
the physical vacuum state, the field $\phi$ has an expectation
value designated by $\phi_{vac}$. Take the state of any single
$i$th particle. Define its coupling $g_i$ to $\phi$ by
$$
m_i=g_i~\phi_{vac},\eqno(16)
$$
where $m_i$ is its physical mass. Next, consider the
transformation
$$
\phi \rightarrow \phi+c,\eqno(17)
$$
where $c$ is a constant; correspondingly the vacuum state is
changed to an excited state, in which the expectation value of
$\phi$ changes accordingly
$$
\phi_{vac} \rightarrow \phi_{vac}+c.\eqno(18)
$$
Hence the mass $m_i$ of any $i$th particle will also be altered,
with
$$
m_i \rightarrow m_i+g_i~c\eqno(19)
$$
(except $\phi$, its mass changes in a different way). Choose the
constant $c$ to be given by
$$
c=-\phi_{vac}.\eqno(20)
$$
Then, (18) and (19) become
$$
\phi_{vac} \rightarrow 0\eqno(21)
$$
and
$$
m_i \rightarrow 0.\eqno(22)
$$

\begin{figure}[h]
 \centerline{
\epsfig{file=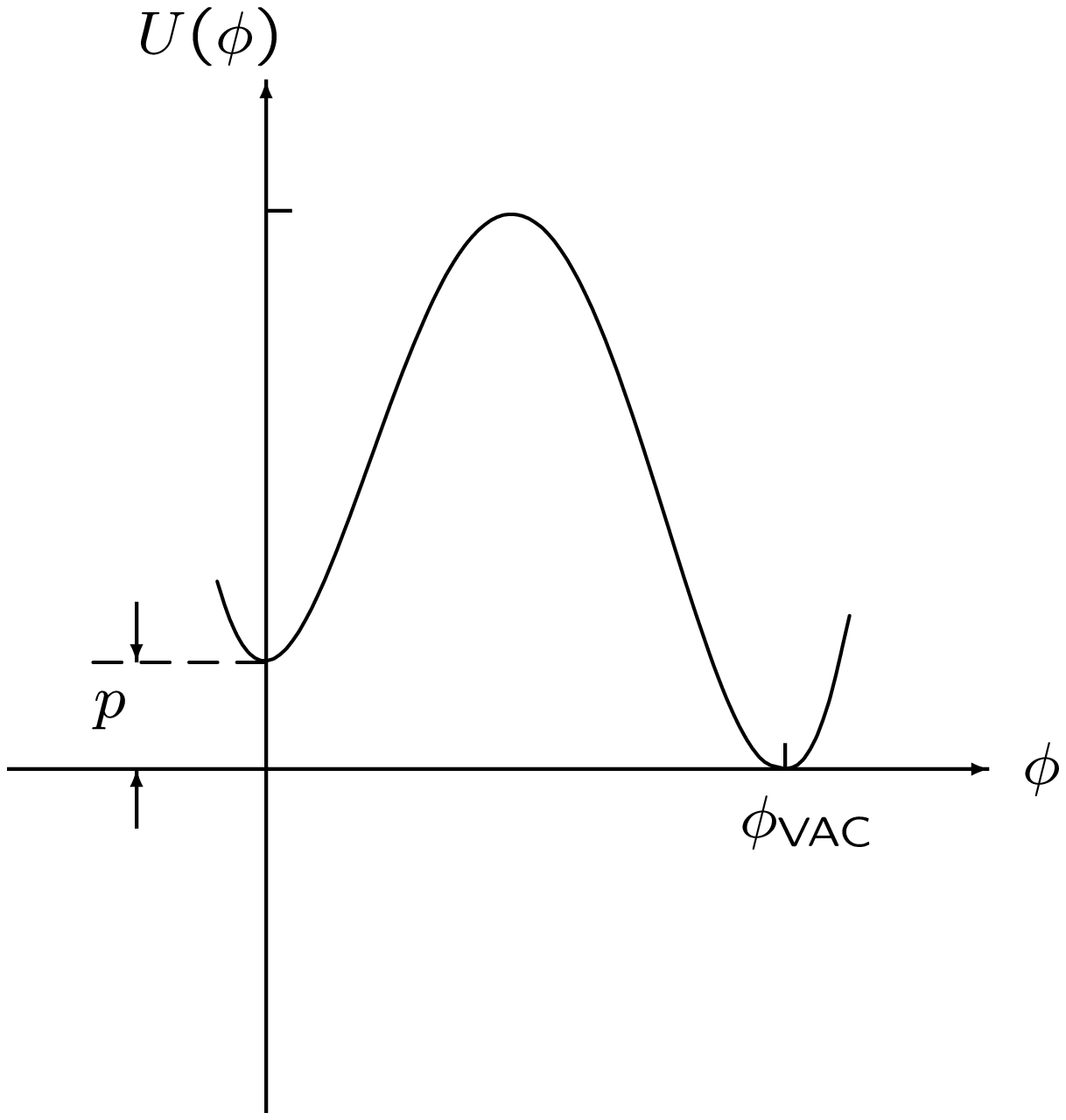, width=8cm, height=6.8cm}}
\vspace{.2cm}
 \centerline{{\normalsize \sf Fig. 2~ A phenomenological potential
 function $U(\phi)$.}}
\end{figure}

Since in any field theory, one can always construct a scalar
component $\phi$, this means for any single physical $i$th
particle, there exists an excited single particle state in which
its mass is zero. The generality of this argument is the origin of
negative pressure, as we shall see.

The transformation (17) transforms the vacuum state to an excited
state with a different expectation value from $\phi_{vac}$. Its
excited energy is proportional to the energy density function
$U(\phi)$ and the volume $\Omega$ of the excitation. An example of
$U(\phi)$ is given in Fig.~2, with the abscissa $\phi$ denoting
the value $\phi_{vac}+c$ in (18).

For a single particle state in an infinite volume, transformations
(21) and (22) lead to excited states with an infinite excitation
energy. But for a system of particles within  finite volume, such
changes may occur spontaneously.

\begin{figure}[h]
 \centerline{
\epsfig{file=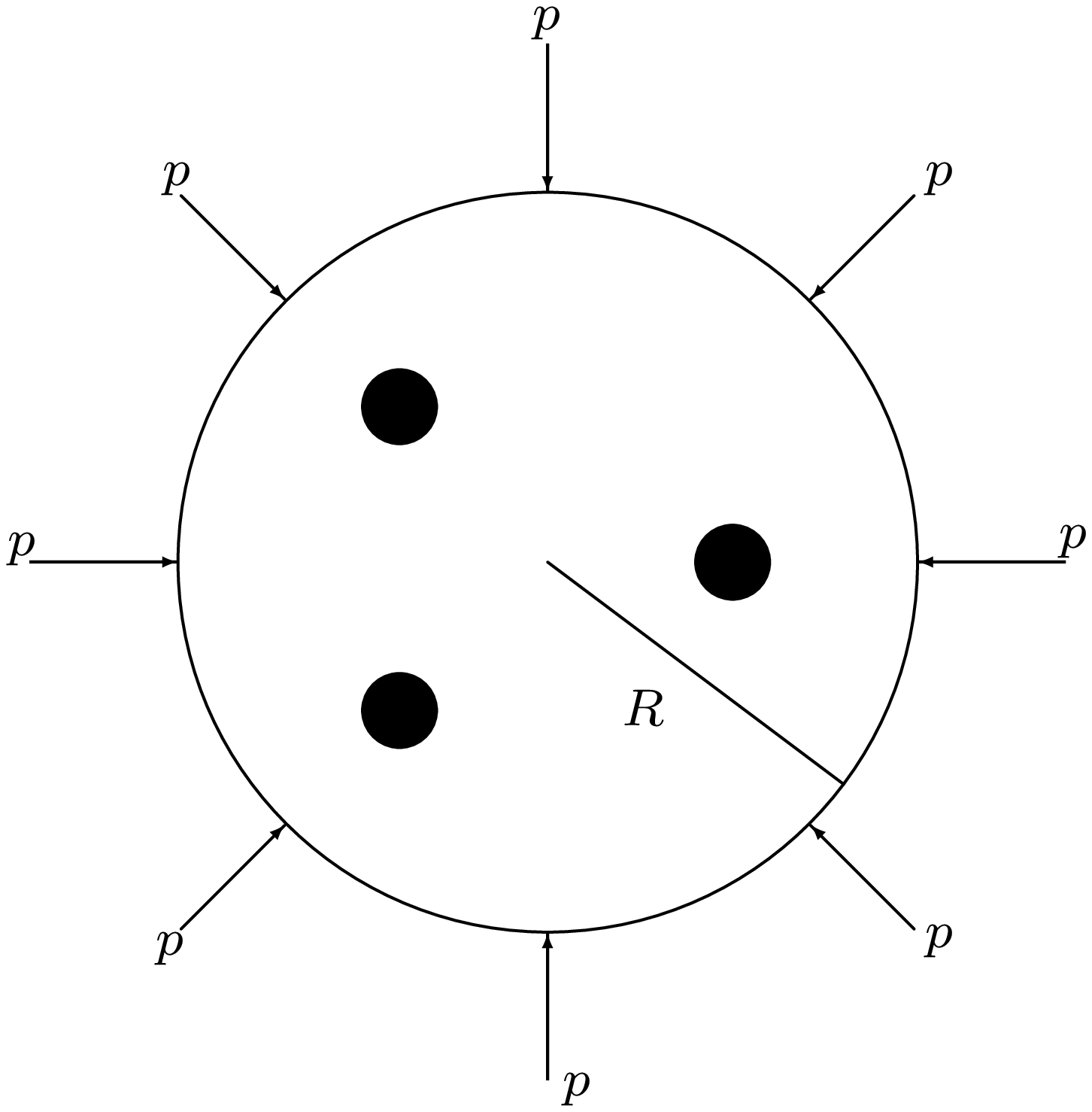, width=8cm, height=6.8cm}} \vspace{.5cm}

{\normalsize \sf
\begin{tabular}{rl}
~~~~~~~~~~~~~~~~~~~~~~~~~~~~~~~~~~~~~~$E-\frac{4\pi}{3}R^3p$&$=N_Q\frac{2.04}{R}$\\
$4\pi R^2p$&$=N_Q\frac{2.04}{R^2}$\\
$\rho_E$&$\equiv \frac{E}{4\pi R^3/3}=4p$\\
~~~~~~~~~~~~~~~~~~~~~~~~~~~$\rho_Q$&$\equiv N_Q\frac{2.04}{4\pi R^4/3}=3p$\\
\end{tabular}
}

\vspace{0.5cm}
 \centerline{{\normalsize \sf Fig. 3~ MIT bag model}}
\end{figure}

Fig.~3 gives the energy balance of the MIT bag model of a single
nucleon, consisting of three quarks. One assumes the expectation
value of $\phi$ to be $\phi_{vac}$ outside the bag, but zero
inside. Thus the quark mass is zero inside the bag, yielding a
quark matter energy $E_Q$ proportional to $1/R$, where $R$ is the
bag radius. Since the magnitude $E_p$ of the energy due to the
negative pressure $-p$ is proportional to $R^3$, from
equipartition of energy it follows that
$$
E_Q=3E_p.\eqno(23)
$$
The total energy $E$ of the system is determined by
$$
E-E_p=E_Q.\eqno(24)
$$
From (23) it follows then
$$
E=4E_p.\eqno(25)
$$

For more complex bag models, the ratio of the total energy $E$ to
the energy $E_p$ due to the negative pressure can vary, but $E_p$
is always a sizable fraction of $E$, like the virial coefficient
in the kinetic theory of gases.

The Higgs field in the electroweak theory is a member of a larger
family of spin $0$ fields which may all be called the inertia
field. Because the inertia field can convert inertia to energy and
vice versa, it connects the microscopic world (like $W^\pm$ and
$Z^0$) to the macroscopic world (like the cosmological constant).
In this view, the origin of all symmetry breaking is due to the
inertia field, and the action of all spin $\neq 0$ fields are
symmetry conserving.

Thus, $W^\pm$ and $Z^0$ are tiny bags, the nucleons are small
bags, the sQGP are manifestations of bigger bags and our whole
universe is a very, very large bag. Through the inertia field, our
microscopic world of particles becomes closely connected to the
macroscopic world of our universe.

After 50 years since the discovery of parity nonconservaion in the
weak interaction, we may poise at the beginning of search for the
origin of all symmetry violations in fundamental physics.

}

\newpage

\section*{\Large \sf References}

{\normalsize \sf

\noindent [1] C. S. Wu (Columbia University) and E. Ambler, R. W.
Hayward,

~D. D. Hoppes and R. P. Hudson (National Bureau of Standards),

~Phys. Rev. 105, 1413(1957)\\
~[2] C. S. Wu, Adventures in Experimental Physics ("Gamma"
volume),

~ed. B. Maglich, Princeton, World Science Communications, 1972,
p102\\
~[3] R. L. Garwin, L. M. Lederman and M. Weinrich, Phys. Rev. 105,
1415(1957);

~J. I. Friedman and V. L. Telegdi, Phys. Rev. 105,
1681(1957)\\
~[4] T. D. Lee and C. N. Yang, Phys. Rev. 104, 254(1956)\\
~[5] W. Pauli, article in Niels Bohr and the Development of
Physics,

~Pergamon Press, London, 1955\\
~[6] K. Lande, E. T. Broth, J. Impeduglia and L. M. Lederman,

~Phys. Rev. 103, 1901(1956)\\
~[7] M. Gell-mann and A. Pais, Phys. Rev. 97, 1387(1955)\\
~[8] T. D. Lee, R. Oehme and C. N. Yang, Phys. Rev. 106,
340(1957)\\
~[9] J. H. Christenson, J. W. Cronin, V. L. Fitch and R. Turlay,

~Phys. Rev. Lett. 13, 138(1964)\\
~[10] N. Cabibbo, Phys. Rev. Lett. 10, 531(1963);

~M. Kobayashi and T. Maskawa, Prog. Theo. Phys. 49, 652(1973)\\
~[11] S. Eidelman et al. (Particle Data Group), Phys. Lett. B592,
1(2004)\\
~[12] R. Friedberg and T. D. Lee (unpublished)\\
~[13] T. D. Lee, Phys. Rev. D8, 1226(1973)\\
~[14] P. W. Higgs, Phys. Lett. 12, 132(1964)\\
~[15] R. N. Mohapatra and J. C. Pati, Phys. Rev. D11, 566(1975)\\
~[16] See also M. S. Chanowitz, Phys. Rev. D66, 073002(2002)\\
~[17] L. N. Cooper, Phys. Rev. 104, 1189(1956); J. Bardeen, L. N.
Cooper

~and J. R. Schrieffer, Phys. Rev. 106, 162(1957); 108,
1175(1957)\\
~[18] J. Schwinger, Ann. Phys. 2, 407(1957)\\
~[19] P. J. E. Peebles and B. Ratra, Rev. Mod. Phys. 75, 559(2003)\\
~[20] P. Schewe and B.Stein, AIP Bulletin of Phys. News, no.757,

~December 7, 2005\\
~[21] T. D. Lee and G. C. Wick, Phys. Rev. D9, 2291(1974);

~T. D. Lee, A possible New Form of Matter at High Density,

~Report of the Workshop on BEV/Nucleon Collisions of Heavy Ions

~~~~~~~---How and Why, Bear Mountain, 1974 (BNL no.50445), 1;

~T. D. Lee, Annual Bevatron Users' Meeting, Lawrence Berkeley
Lab., 1974

~(Rev. Mod. Phys. 47, 267(1975))\\
~[22] A. Chodos, R. J. Jaffe, K. Johnson, C. B. Thorn and V. F.
Weiskopf,

~Phys. Rev. D9, 3471(1974)\\
~[23] D. A. M. Dirac, Proc. R. Soc., A268, 57(1962).\\

}

\section*{\Large \sf Appendix A}

\begin{center}
2006 APS April Meeting\\
Saturday-Tuesday, April 22-25, 2006, Dallas, TX
\end{center}

Session E5a: 50 Years Since the discovery of Parity
Nonconservation

~~~~~~~~~~~~~~~~~in the Weak Interaction {\large \rm I}\\

{\normalsize \sf

Sponsoring Units: FHP DPF

Chair: Natalie Roe, Lawrence Berkeley National Laboratory

Hyatt Regency Dallas-Pegasus B\\

\begin{tabular}{ll}
  Saturday, April 22, 2006& \underline{E5a.00001: New Insights to Old Problems}\\
  3:30pm~-~3:57pm & Invited Speaker: Tsung Dao Lee, Columbia University\\
\end{tabular}
\begin{center}
 \underline{Abstract}\\
\end{center}
From the history of the $\theta -\tau$ puzzle and the discovery of
parity non-conservation in 1956, we review the current status of
discrete symmetry violations in the weak interaction. Possible
origins of these symmetry violations are discussed.\\

\begin{tabular}{ll}
  Saturday, April 22, 2006 &\underline{E5a.00002: Cracks in the Mirror: Saga of a 36 Hour Experiment}  \\
  3:57pm~-~4:24pm & Invited Speaker: Leon Lederman\\
                  &~~~~~~Fermi National Accelerator Laboratory  \\
\end{tabular}
\begin{center}
     \underline{Abstract}  \\
\end{center}
The history of the fall of parity, mirror symmetry, emerges from a
puzzle in the behavior of particles ($\tau-\theta$ puzzle). This
stimulated the Lee-Yang paper of mid-1956 questioning the validity
of parity in the weak interactions. They specifically raised the
issue of the weak decays $\pi^{\pm} \rightarrow \mu^\pm+\nu$  and
$\mu^{\pm}\rightarrow e^\pm+2\nu$ . In subsequent detailed
discussions between C.S. Wu and T.D. Lee, Wu designed a
collaborative experiment with physicists from the Bureau of
Standards in Washington D.C., which examined the decay of
Co$^{60}$, an easily polarizable nucleus. Early positive results
of the Wu experiment were discussed at a Friday lunch
traditionally "chaired" by T.D. Lee. The precise date was the
Friday of the first working week after the New Year, 1957. Here,
for what was probably the first time, the possibility was raised
that the failure of parity conservation could be a large effect.
The conversation at the very traditional Chinese lunch was
exciting. This new concept stirred me in my drive from Columbia to
home in Irvington, actually a short walk to the Nevis laboratory
where Columbia's 400 MeV synchrocyclotron lab. was housed. The
events of the next few days are the substance of my paper. By
Tuesday noon, the word had spread around the world that parity
conservation was dead. By that time we had a 20 $\sigma$ effect
and many of the essential tests of validity of our experiment were
done. Some of the consequences important to that time, and some
still relevant in 2006 will be presented.\\

\newpage

\begin{tabular}{ll}
  Saturday, April 22, 2006 &\underline{E5a.00003: Do Left-handed Neutrino Have Rights Too?}  \\
  4:24pm~-~4:51pm & Invited Speaker: Janet Conrad, Columbia University  \\
 \end{tabular}
\begin{center}
\underline{Abstract}  \\
\end{center}
As with any great discovery, the result of C.S. Wu's experiment on
parity violation opened up many more questions. This talk explores
these questions in light of recent discoveries about neutrino
properties and the potential
 for new discoveries at the highest energy scales.\\

 \vspace{0.5cm}

{\large \sf

Session E5b: 50 Years Since the discovery of Parity
Nonconservation

~~~~~~~~~~~~~~~~~in the Weak Interaction {\large \rm II}\\

}

{\normalsize \sf

Sponsoring Units: FHP DPF

Chair: Noemie Koller, Rutgers University

Hyatt Regency Dallas-Pegasus B\\

\begin{tabular}{ll}
  Saturday, April 22, 2006 &\underline{E5b.00001: In Which Direction Is the Door?}  \\
  5:00pm~-~5:27pm & Invited Speaker: C. N. Yang, SUNY-Stony Brook, Emeritus  \\
 \end{tabular}
\begin{center}
  \underline{Abstract}  \\
 This abstract was not received electronically.\\
\end{center}


{\large \sf
 Session P10: 50 Years Since the discovery of Parity
Nonconservation

~~~~~~~~~~~~~~~~~in the Weak Interaction {\large \rm III}\\

}

{\normalsize \sf

Sponsoring Units: FHP DNP

Chair: Susan Seestrom, Los Alamos National Laboratory

Hyatt Regency Dallas-Cunnberland C\\

\begin{tabular}{ll}
Monday, April 24, 2006& \underline{P10.00001: The Question Answered}\\
  10:45~-~11:21am & Invited speaker: Ralph P. Hudson\\
                  &~~~~~~National Institute of Standards and Technology (retired)\\
\end{tabular}
\begin{center}
\underline{Abstract}  \\
\end{center}
Fifty years ago, theoreticians T.D. Lee and C.N. Yang reached the
conclusion that no experimental evidence existed to show that
parity is conserved in weak interactions. The speaker will explain
how it happened that a team of physicists at the U.S. National
Bureau of Standards in Washington D.C. came to perform the
revolutionary experiment that demonstrated the failure of parity
conservation in
nuclear beta-decay.\\

\newpage

\begin{tabular}{ll}
  Monday, April 24, 2006 &\underline{P10.00002: Chien-Shiung Wu as a Person and a Scientist}  \\
  11:21am~-~11:57am & Invited Speaker: Vincent Yuan, Los Alamos National Laboratory  \\
\end{tabular}
\begin{center}
     \underline{Abstract}  \\
\end{center}
Chien-Shiung Wu is well known for her outstanding contributions to
Nuclear Physics in Parity Violation and Beta Decay. I will discuss
my view of her as seen from my vantage point as her son.\\

\begin{tabular}{ll}
  Monday, April 224, 2006 &\underline{P10.00003: Looking through the Mirror: Future Directions }  \\
  11:57am~-~12:33am & \underline{in Parity Violations}\\
  & Invited Speaker: M. Ramsay-Musolf\\
  &~~~~~~California Institute of Technology  \\
\end{tabular}
\begin{center}
    \underline{Abstract}  \\
\end{center}
Over the past fifty years, studies of parity violation (PV)
involving atoms, nuclei, and elementary particles have taught us a
great deal about the electroweak and strong interactions. The
future of this field promises to be equally rich. In this talk, I
discuss new initiatives using PV to study weak interactions among
quarks, to probe the structure of the nucleon, and to search for
physics beyond the Standard Model.

\end{document}